\documentclass[preprint,prb,aps,titlepage,byrevtex,floatfix]{revtex4}
\usepackage{epsfig}
\usepackage{rotating}
\usepackage{color}
\begin{document}
\title{The freezing tendency towards 4-coordinated amorphous network causes
increase in heat capacity of supercooled Stillinger-Weber silicon}
\author{Pankaj A. Apte, Nandlal Pingua, Arvind Kumar Gautam, Uday Kumar}
\affiliation{Department of Chemical Engineering,
         Indian Institute of Technology Kanpur,
         Kanpur,
         U.P, India  208016}
\author{Soohaeng Yoo Willow}
\affiliation{Department of Chemistry,
        University of Illinois, Urbana--Champaign, 600 South Mathews
        Avenue, Urbana, Illinois 61801, USA}
\author{Xiao Cheng Zeng}
\affiliation{Department of Chemistry,
         University of Nebraska-Lincoln, Lincoln, Nebraska 68588, USA}
\author{B. D. Kulkarni}
\affiliation{CSIR-National Chemical Laboratory, Pashan Road, Pune 411008, India}
\vspace{20ex}
\date{{Submitted} to a Peer-reviewed Journal on \underline{28 January 2014}}
\begin{abstract}
The supercooled liquid silicon, modeled by Stillinger-Weber potential, shows
anomalous increase in heat capacity $C_p$, with a maximum $C_p$ value close 
to 1060 K at zero pressure. 
We study equilibration and relaxation of the supercooled SW Si,
in the temperature range of 1060 K--1070 K at 
zero pressure.  
We find that as the relaxation of the metastable supercooled liquid phase 
initiates, a straight line 
region (SLR) is formed in cumulative potential energy distributions.  The configurational
temperature corresponding to the SLR is close to 1060 K, which was earlier
identified as the freezing temperature of 4-coordinated amorphous network.
The SLR is found to be tangential to the distribution of the metastable liquid phase
and thus influences the broadness of the distribution.  As the
bath temperature is reduced from 1070 K to 1060 K, the effective temperature approaches
the bath temperature which results in broadening of the metastable phase distribution.
This, in turn, causes an increase in 
overall fluctuations of potential energy and hence an increase of heat capacity.
We also find that during initial stages of relaxation, 4-coordinated atoms form 
6-membered rings with a chair--like structure and other structural units that
indicate crystallization.   Simultaneously a strong correlation
is established between the number of chair-shaped 6-membered rings 
and the number of 4-coordinated atoms
in the system.  This shows that all properties related to 4-coordinated particles
are highly correlated as the SLR is formed in potential energy distributions
and this can be interpreted as a consequence 
of `freezing' of amorphous network formed by 4-coordinated particles.  
\end{abstract}
\maketitle
\section{INTRODUCTION}
The supercooled (metastable) liquid silicon, modeled by Stillinger--Weber (SW) potential model,~\cite{STILLINGER85}
exhibits an increase in constant pressure heat capacity $C_p$ upon deeper supercooling, with 
a maximum $C_p$ value close to 1060 K at zero pressure.~\cite{HUJO11}  
The reason
for this anomalous behavior is not fully understood.  
In a previous work,~\cite{APTE12} the properties
of metastable equilibrium liquid of SW-Si, 
in the temperature range of 1060--1070 K,
were ascertained by means of free energy
computations.  
Here we analyze the properties of the metastable equilibrium liquid in more detail vis-a-vis
the nature of the relaxation process. 
Currently, there is a debate~\cite{HUJO11,LIMMER11,LIMMER13}
on whether there is a first-order transition~\cite{SASTRY03,VASISHT11} 
from the high-density liquid phase 
to a metastable equilibrium low density liquid phase near 1060 K.  
However in this work, our
focus is mainly placed on the properties of metastable SW-Si liquid phase in the temperature range of 1060 - 1070 K only and 
we are only interested in the initial 
stages of relaxation of the metastable SW-Si liquid phase.  

For understanding the nature of the relaxation, we focus particularly on
a few previous studies as follows.  In a detailed analysis of structure
of amorphous phase of SW-Si, Luedtke and Landmann~\cite{LUEDTKE89} found
1061 K as the ``effective temperature'' (denoted as $T_4^*$),
below which amorphous network
`freezes' in SW-Si.   This conclusion was based on analysis of the straight
line region (SLR) in potential energy distributions of 4-coordinated 
particles (see Fig.~16 of Ref.~\onlinecite{LUEDTKE89})
We interpret the ``freezing" of amorphous network at $1061$ K or below to mean that
there is a reduction in mobility
of 4-coordinated particles. Indeed, Vasisht et al. \onlinecite{VASISHT11} 
reported significant reduction in diffusivity ($< 10^{-8}$ cm$^2$/s) for local structure of SW-Si with coordination number of $\sim$4 at 1070 K. 
Note also that Luedtke and Landman
had mentioned that ``five fold coordinated atoms maintain their
mobility down to $T_5^*$'' (page 1173 of Ref.~\onlinecite{LUEDTKE89})
and $T_5^*$ was found to be 806 K.   
We also note that the appearance of SLR 
in potential energy distributions has been found in numerical simulations
of foam system under shear (see Fig. 4 of Ref.~\onlinecite{ONO02}).
Based on the following two statements, it appears that 
the relaxation of metastable liquid phase can also be termed as 
a jamming transition: 
(i) ``Indeed we note that the calculated value of $T_4^*$ is close to the temperature at
which the transition from liquid to amorphous silicon commences during our slow quench"
(page 1173 of Ref.~\onlinecite{LUEDTKE89}) and (
ii) ``... This suggests that the
concept of effective temperature should be useful for {\it any} system near
the onset of jamming" (page 095703-4 of Ref.~\onlinecite{ONO02}).  There are several other
studies that point to jamming transition near 1060 K and zero pressure.
Sastry and Angell~\cite{SASTRY03} have noted that there is
a reduction in diffusivity by two orders of magnitude when the high temperature
liquid phase
undergoes a transition to a low temperature liquid phase near 1060 K, and in a later paper
a sudden decrease of heat capacity 
across the same transition near 1060 K has been reported~(Fig.~1 of Ref.\onlinecite{JAKSE09}).
The reduced mobility of 4-coordinated particles would cause both the diffusivity and
the heat capacity to decrease. 
This also
is consistent with the above conclusion that the relaxation of metastable liquid phase involves
freezing or jamming.  

The focus of our present work is how the freezing of 4-coordinated particles
affects potential energy distributions
and relaxation of metastable equilibrium liquid phase, when the external (bath)
temperature is in the range of 1060-1070 K.  
In MC simulations, the kinetic
signature of freezing (or jamming), i.e., reduction in mobility 
cannot be detected, but
the statistical mechanical 
signature~\cite{LUEDTKE89} i.e., appearance of SLR in the potential
energy distribution can be detected, and we study this latter aspect of 
the metastable liquid phase relaxation.
In the title of the paper we have used the word freezing ``tendency" to
mean the onset of jamming transition which is signaled by 
attainment of tangential condition, i.e., 
complete equilibration of the metastable liquid phase and the onset of
relaxation (or jamming) of the system
is achieved when the SLR is tangential to the
potential energy distribution of the
metastable liquid phase.  We conclude that this tangential condition 
causes
the increase in $C_p$ of metastable liquid phase.  
The details of our methodology and analysis 
are given in the following sections.

\section{Identification of metastable equilibrium states}

The partition function associated with the metastable
equilibrium liquid is given by
\begin{equation}
    Y  =  \frac{1}{h^{3N} N!} \int~dV  \exp(-\beta p V)
          \int~d {\bf p}^N \exp[-\beta \sum_{i=1}^{N} \frac{p_i^2}{2m}] 
          \int~d {\bf r}^N \exp[-\beta \Phi({\bf r}^N)] 
\label{eq:partition1}
\end{equation}
where 
$V$ is the total volume, 
${\bf r}^N = ({\bf r}_1, \cdots, {\bf r}_N)$ is the set of particle coordinates, 
${\bf p}^N = ({\bf p}_1, \cdots, {\bf p}_N)$ is the set of momenta of particles, and
$\Phi({\bf r}^N)$ is the total potential energy.  
To define Y, it is important to identify the limits on the various integrals that appear in 
Eq.~(\ref{eq:partition1}).  These are determined by the region of the phase space which is explored
by the metastable phase.  As is normal practice,
it is possible to integrate over the momenta of the particles analytically 
by extending the limits from
$-\infty$ to $+\infty$.  
This extension of the limits does not cause significant error,
since the exponential term containing $-p^2$ decreases sufficiently with the small deviations
of $p$ explored by the system with respect to the equilibrium (average) value.~\cite{LANDAU-5}  
Integrating over the momenta, we get the following form of the partition function:
\begin{equation}
    Y  =  \frac{1}{\Lambda^{3N} N!} \int~dV \exp(- \beta p V) 
                   \int~d {\bf r}^N \exp[-\beta \Phi({\bf r}^N)],
\label{eq:partition2}
\end{equation}
where $\Lambda$ is the thermal wavelength.

Any metastable phase, after a sufficient time, undergoes relaxation process 
towards the stable phase. 
However,  as mentioned in Ref.~\onlinecite{LANDAU-5},
this relaxation process is not considered in computing the average 
properties of the metastable phase. 
Hence we consider the phase space which the metastable phase
explores {\it before} the relaxation (or crystallization
 in this study) begins, to define the limits over the
integrals in the partition function.  Limmer and Chandler~\cite{LIMMER11,LIMMER13}
recently showed that the metastable liquid of SW-Si undergoes a partial crystallization process in the supercooled region.
Hence we consider formation of clusters of 4-coordinated particles that 
indicate crystallization,~\cite{LUEDTKE89,BEAUCAGE05}
in order to detect the initiation of the relaxation process.

According to thermodynamics, 
the metastable equilibrium state must be associated with
a unique free energy minimum.  
Recent studies by Limmer and Chandler~\cite{LIMMER11,LIMMER13} 
(on supercooled SW-Si) 
underscore the necessity 
of  identifying a free energy minimum associated with the metastable equilibrium liquid.
Thus all the micro-states  
associated with the unique free energy minimum must be considered in computing metastable
phase properties.   Since the metastable liquid phase of SW-Si 
in the NPT ensemble, is thermodynamically contiguous 
with the stable liquid above the melting temperature,
there must be an associated free energy
minimum with respect to fluctuating variables in NPT ensemble
namely volume and energy.  Thus in NPT-MC trajectories, 
we {\it assume} that
the point beyond which the 
free energy minimum with respect to volume and energy
is not accessible as the initiation point 
of the relaxation process.  
We check this assumption by computing the structural 
changes associated with 4-coordinated particles. 

\section{Results and Discussion}

In what follows, we study the equilibration and relaxation of the 
metastable liquid phase of SW-Si
in the temperature range of 1060--1070 K 
at zero pressure using isothermal--isobaric (NPT) Monte Carlo (MC)
simulations.  We use cubic simulation box with periodic boundary conditions and two different
system sizes of 512 and 4096 particles.  
Table I of Ref.~\onlinecite{APTE12} provided the
the equilibrium energies and densities of the metastable liquid phases
on the basis of the fact that these properties, 
together with the corresponding excess
Gibbs free energies, 
follow the Gibbs Helmholtz equation.   
Figures~\ref{fig:t1060} and ~\ref{fig:str1060} show the NPT-MC trajectory at 1060 K with $N~=~512$
particles, along with block averages of the clusters of 4-coordinated particles
$N_{r6}$, $N_{r6c}$, $N_{d10}$, and $N_{w12}$ formed along the trajectory.  
This is the same trajectory
as in Figs.~1 and 2 of Ref.~\onlinecite{APTE12}.  
As mentioned above, the initiation point of the 
relaxation process (point R in these figures) is the point beyond which the free energy
minimum is no longer accessed. 

Our procedure to locate the R-point is as follows. 
Based on the NPT-MC trajectory, we first compute 
$N_c(\phi,\rho)$, which  is the number of micro-states
along the trajectory with per particle potential energy and density in the interval
$(\phi,\rho)$ and $(\phi+\Delta \phi, \rho+\Delta \rho)$.  Note that density $\rho = N/V$ can 
be used
in place of the volume $V$ in Eq.~(\ref{eq:partition2}), simply by change of variables.
If $p(\phi,\rho)$ is the distribution
function, then $N_c \propto p(\phi,\rho) \Delta \phi \Delta \rho$.
Hence $\log N_c = \log p(\phi,\rho) + \mbox{constant}$, where $\Delta \phi$ and $\Delta \rho$ 
are the
widths of the bins and are absorbed in the constant term in the above equation.  Since
$-\beta G (\phi,\rho) = \log p(\phi,\rho)$, we get the relation 
$\log N_c (\phi,\rho) = -\beta G (\phi,\rho)+ \mbox{constant}$
for the free energy surface.  
Based on the {\it final} free energy surface thus obtained from $N_c$,
we identify the free energy minimum associated
with the metastable liquid.  By {\it final}, we mean that the portion of the
free energy surface corresponding to the metastable liquid
is not likely to be visited again by the system.  
We then consider the micro-states visited along the NPT-MC trajectory in which
this free energy minimum is accessed (see Figs.~\ref{fig:t1060} and \ref{fig:str1060}).
The point beyond which ($\phi_m, \rho_m$) is not accessible
is the R-point and is assumed to be the initiation point of the relaxation (i.e. crystallization) 
process. 
The average properties of the metastable equilibrium liquid are
computed based on the trajectory upto point R.  Thus average per particle potential energy
(for the 1060 K trajectory in Figs.~\ref{fig:t1060} and ~\ref{fig:str1060}) 
is found to be
$\langle \phi \rangle = -1.8272$ and 
$\langle \rho \rangle = 0.4739$.  
These values are very close to that reported in Table I of
Ref.~\onlinecite{APTE12}.   Note that in Ref.~\onlinecite{APTE12}, 
the relaxation point was detected as the point at which
cumulative average potential energy shows a continuous decrease.  

To test whether relaxation to the stable crystal phase (i.e., crystallization) 
indeed initiates at the R-point, we identify formation of structural units
of 4-coordinated particles that indicate crystallization.~\cite{LUEDTKE89,ZALLEN83,BEAUCAGE05}
In Figs.~\ref{fig:t1060} and \ref{fig:str1060}, we have shown the number 
of 6-member rings $N_{r6}$,  
6-member rings with chair--like shape $N_{r6c}$
(similar to that in stable crystal phase), 
d-10 clusters~\cite{BEAUCAGE05} $N_{d10}$ i.e., clusters containing a tetrahedral unit connected to a six-member ring with
the same connectivity as in diamond unit cell, and w-12 clusters~\cite{BEAUCAGE05} $N_{w12}$, 
i.e., clusters containing two 
6-member rings connected at alternate positions as in the Wurtzite unit cell.  The connectivity
in d-10 and w-12 clusters is taken to be the same as that shown in Fig.~1 of Ref.~\onlinecite{BEAUCAGE05}.  
Though $N_{r6}$, $N_{r6c}$, $N_{d10}$, and $N_{w12}$ show wide fluctuations
(see Figs.~\ref{fig:t1060} and \ref{fig:str1060})
after the relaxation
point (R-point), these quantities appear to be correlated.  
It can also be seen that before the relaxation,
only the six-member rings $N_{r6}$ are present in appreciable numbers, 
while $N_{r6c}$, $N_{d10}$,
and $N_{w12}$ are negligible in comparison.  
In Fig.~\ref{fig:n4r6c1060}, we plot block averages of the fraction of  
4-coordinated particles $f_4 = (N_4/N)$, where $N=512$ is the total number of particles.  
In the same figure, we have also plotted
$\langle \Delta N_4 \Delta N_{r6c} \rangle_b$ (please refer to ordinate on the right hand side
in Fig.~\ref{fig:n4r6c1060}), 
which is a measure of correlation between the fluctuations
of the number of 4-coordinated atoms $N_4$ 
and the number of chair--like 6-member rings $N_{r6c}$.  
Before the relaxation,
the quantity 
$\langle \Delta N_4 \Delta N_{r6c} \rangle_b$
fluctuates close to zero value in the metastable equilibrium liquid, 
indicating a weak correlation.  However after the relaxation,
$\langle \Delta N_4 \Delta N_{r6c} \rangle_b$
shows relatively large deviation from zero value, indicating 
strong correlation.  It is to be noted
that $N_4$ is the number of all 4-coordinated particles, which includes 
those 4-coordinated particles which are not part 
of chair--like six membered rings.  Therefore, we
expect that fluctuations in
all quantities related configuration 
of 4-coordinated particles 
will show strong correlation
after the relaxation initiates.  
This indicates that amorphous network formed
by 4-coordinated particles is `frozen' after the R-point,
which is due to significant reduction in mobility~\cite{LUEDTKE89} 
of  4-coordinated particles.

In Fig.~\ref{fig:config6_1}, one can see that six member rings are present in the metastable
liquid before the relaxation, 
but do not have
a regular shape.  However, after the relaxation, the six member rings in chair--like shape are
formed as seen in Fig.~\ref{fig:config6_6}.  We also note that 6-member rings with 
boat--like shape~\cite{ZALLEN83}
are also formed (though not visible in Fig.~\ref{fig:config6_6})
after the relaxation, but such rings are not present in the stable crystal phase.    
To identify 6-membered rings with a chair--like shape, we note that~\cite{ZALLEN83}
all 6-bonds in a chair-shaped rings are in a staggered
configuration,  which corresponds to the dihedral angle of   
$\varphi = 60^0$.  
In the stable crystal phase, all 6-membered rings are in a chair-like 
configuration~\cite{ZALLEN83}.
Hence, in order to identify rings with approximately chair--like shape, we first studied
the fluctuations of the  average dihedral angle of all 6 bonds forming the rings in the
crystal phase at 1060 K and found that the angle fluctuates be in
the range of $45^0 \le \varphi \le 60^0$.  Thus to identify
the six-membered rings with a chair-like shape : we impose the condition that each of the six
bonds forming the ring the average dihedral angle must be in the above range. 
The dihedral angle associated with a bond joining four coordinated atoms is computed as described
below.~\cite{LUEDTKE89,ZALLEN83} We consider the bond joining
a pair of four coordinated particles (say particle i and particle j).  Excluding the ij bond which
is the common, each atom i and j has three bonds.  We consider a projection of these bonds on a plane
perpendicular and bisecting the ij bond.  
For a given i bond projection, we consider the angle made with the nearest
projection of j-bond (i.e., minimum angle) as the dihedral angle.  Thus for a given pair i and j 
of 4-coordinated
atoms forming the ij bond, there
are three dihedral angles and we then take the average over these three angles which we denote
as $\varphi$ and assign this average dihedral angle  to the ij bond.  For a perfect crystal configuration,
the angle $\varphi$ is exactly $60^0$ for each bond and this is known as the staggered configuration.
We expect that as the liquid undergoes relaxation, the bonds will approach a staggered configuration
similar to the crystal.  

We also find that a straight line region (SLR) is formed in the cumulative 
potential energy distributions during
the initial stages of relaxation as seen in Fig.~\ref{fig:fig1}.  A straight
line is fitted to the SLR with a correlation coefficient of
a correlation coefficient 
of $R^2=99.906\%$ and a configurational temperature [obtained as explained
in Eq.~(\ref{eq:te2}) below] of $T_c = 1056.53$ K.   Note also the similarity
of the SLR in Fig.~\ref{fig:fig1} and that in Fig.~4 (a) of Ref.~\onlinecite{ONO02}.  
Interestingly, we also find
that the SLR is, at least approximately, 
tangential to the distribution corresponding to the metastable equilibrium metastable liquid phase.
As will be seen below, this tangential condition is always attained by the metastable liquid phase
distribution. This leads us to the important conclusion 
that complete equilibration of the metastable phase is achieved and  
the relaxation initiates when the distribution corresponding to the metastable liquid phase
is tangential to 
the SLR that eventually appears.  
The significance of the SLR in the potential energy 
distributions can be explained as follows.
The isothermal--isobaric partition function in Eq.~(\ref{eq:partition2})
at zero pressure can be written as,
\begin{eqnarray} 
     Y  &=& (\mbox{constant}) 
           \int d V \int d\Phi~\exp[-\beta \phi+S(\phi,V)/k_B] \\ \nonumber
        &=& (\mbox{constant}) 
            \int d\Phi~~\exp(-\beta \Phi) \int dV~~\exp[S(\Phi,V)/k_B] \\ \nonumber
        &=& (\mbox{constant}) 
            \int d\Phi~~\exp(-\beta \Phi)~~\exp[S(\Phi)/k_B] \\ \nonumber
        &=& (\mbox{constant}) 
           \int d\Phi~~p(\Phi) \\ \nonumber
\label{eq:partition3}
\end{eqnarray}
where  
$\exp[S(V, \Phi)/k_B]$ is the number of all possible configurations with a potential
energy between $\Phi$ and $\Phi + d\Phi$ and volume $V$.   After integrating
over the volume, the we get expression for $Y$ in terms of 
$\exp[S(\Phi)/k_B]$, which is the number of configurations corresponding to the
metastable phase for all value of $V$.   It is also clear that 
$\log p(\Phi) = -\beta \Phi + S(\Phi)/k_B$.
We define the {\it configurational} temperature as 
\begin{equation}
     \frac{1}{T_c} = \frac{d S}{d \Phi}
\label{eq:te1}
\end{equation}
Thus the configurational temperature can be computed from $\log p(\Phi)$ as follows :
\begin{equation}
     \frac{1}{k_B T_c} = \frac{d \log p(\Phi)}{d \Phi}+ \beta
\label{eq:te2}
\end{equation}
where $\beta = 1/k_B T$ is the reciprocal of the bath temperature $T$.
Ono et. al.~\cite{ONO02} plotted configurational
entropy as a function of potential energy (form a foam system under shear),
which showed SLRs on the lower energy side (see Fig.~4(a) of Ref.~\onlinecite{ONO02}).  
These authors
obtained the effective
temperature by differentiating configurational entropy with respect to
potential energy in the SLR region.  
It is also to be noted that Luedtke and Landman~\cite{LUEDTKE89}
found $1061$ K as the effective temperature for freezing of 4-coordinated 
amorphous network.  This analysis was also based on distribution of per particle potential
energy distributions in the amorphous phase prepared by slow cooling in MD simulations.
Based on these studies, we consider the
the value of $T_c$ corresponding to the SLR as effective
temperature for the system.  

We find that we could generate independent trajectories with system sizes of 512 and 4096
particles  with average potential energies and average densities 
(calculated using the above protocol)
close to the equilibrium values listed in Table I  
of Ref.~\onlinecite{APTE12}.  
In Figs.
~\ref{fig:t1070_N512_cp1300}
--
~\ref{fig:n4r6c1070}, we show the structural changes along the 
trajectory obtained at 1070 K with 512 
particle that shows
effective temperatures very close to 1060 K as seen in Fig~\ref{fig:fig3}.  
We found that the NPT-MC trajectories at 1065 K and 1070 K 
in Ref.~\onlinecite{APTE12} yield a slightly lower
effective temperature of about 1056 K.  
The average properties of the metastable equilibrium 
liquid at 1070 K in Ref.~\onlinecite{APTE12}
 agree closely with the current trajectory, i.e., average
properties of metastable liquid are not found to vary significantly
when effective temperature changes from 1056 K to 1060 K.   
Comparing 
Figs.~\ref{fig:t1070_N512_cp1300}
--
~\ref{fig:n4r6c1070} at 1070 K, with 
the corresponding Figs.~\ref{fig:t1060}
--
~\ref{fig:n4r6c1060} at 1060 K, it can be seen that
the features of the relaxation across the R-point 
are qualitatively similar to that described above for 1060 K trajectory. 
With respect to the formation of SLR (see Fig.~\ref{fig:fig3}), we find that
$T_c = 1059.92$ K which is very close to 1060 K.  Also the mid-point of the SLR
is $\phi_{\mbox{mid}}= -1.827$ which is close the equilibrium (average) value
of per particle potential energy of the metastable liquid phase at 1060 K.  
Most importantly,
the SLR formed in the initial stages of relaxation is,
at least approximately, tangential to the metastable liquid phase distribution 
in Fig.~\ref{fig:fig3}. 
  
In supplementary
information, we have shown 
data for the NPT MC trajectories generated using 4096 particles
at 1060 K and 1065 K [see Figs.~S1--S8 in supplementary information] 
that show average properties
close to the equilibrium values listed in Table I of Ref.~\onlinecite{APTE12}.  
We find that at the relaxation point (R) of 1060 K trajectory in Figs.~S1 and S2, 
the cumulative average
potential energy and density are $-1.8270$ and $0.47398$, respectively.  These values
agree well with the equilibrium values obtained using 512 particle simulations (see Table I
of Ref.~\onlinecite{APTE12}).  However, unlike the 1060 K trajectory with 512 particles,
the quantities  
($N_{r6c}$,$N_{d10}$, and
$N_{w12}$)  
begin to rise slowly across the R-point as seen in Figs.~S1 and S2.  Thus, it cannot
be concluded on the basis of these quantities that R-point corresponds to the initiation
of the relaxation process.  However,   
$\langle \Delta N_4 \Delta N_{r6c} \rangle_b$ rises
sharply at the R-point indicating the initiation of the relaxation process.
The SLR region is formed upon relaxation with an effective
temperature of $T_c = 1056$ K, which is seen in Fig.~S4.  
For 1065 K trajectory with
4096 particles shown in Figs.~S5 and S6, 
we find the cumulative average value of potential energy
and density as $\langle \phi \rangle = -1.822$ and $\langle \rho \rangle = 0.477$,
respectively (which is close to equilibrium values listed in Table I of Ref.~\onlinecite{APTE12}).
As in the case of 1060 K trajectory with 4096 particles, the quantities  
the quantities  
($N_{r6c}$,$N_{d10}$, and
$N_{w12}$) 
increase slowly across the R-point (Figs.~S5 and S6), but there is a sharp increase in
$\langle \Delta N_4 \Delta N_{r6c} \rangle_b$ (see Fig.~S7) which signals beginning
of the relaxation at R-point.  The SLR developed (see Fig.~S8) 
corresponds to a configurational (or
effective) temperature of $T_c = 1059.9$ K and is centered at $\phi_{\mbox{mid}}= -1.827$.
Interestingly, 
this value is close to equilibrium per particle potential energy of metastable liquid phase at 1060 K.
It can also be seen that the complete equilibration of the metastable liquid phase is achieved and the
relaxation initiates when 
the metastable liquid phase distribution is tangential to the SLR in the cumulative potential
energy distributions
obtained for 4096-particle trajectories at 1060 K and 1065 K (see Figs.~S4 and S8).  This
is consistent with the observation for 512 particle trajectories.  

\section{SUMMARY}
In this work, we obtained average (equilibrium) properties of the metastable liquid phase in NPT-MC trajectories
using 512 and 4096 particles, 
based on the thermodynamic requirement that all the micro-states 
associated with a free energy minimum must be considered.  We considered the point (R-point)
beyond which free energy minimum is not accessible as the initiation point for the
relaxation process.   
We found that in 512 particle simulations $N_{r6c}$,
$N_{d10}$, $N_{w12}$ and the magnitude of the correlation 
$\langle \Delta N_4 \Delta N_{r6c} \rangle_b$ increases sharply
after the R-point indicating initiation of the relaxation process.  
The data shows that after the relaxation, a strong
correlation is established between structural quantities related to 4-coordinated
particles. The strong correlation 
indicates that 4-coordinated amorphous network is frozen.  
In case of 4096 particle simulations, as relaxation occurs
($N_{r6c}$,$N_{d10}$, and
$N_{w12}$)  
begin to rise slowly.    
However, we found that the magnitude of the  
correlation between $N_4$ and $N_{r6c}$, i.e., 
$\langle \Delta N_4 \Delta N_{r6c} \rangle_b$ 
changes sharply at the R-point.  This again confirms
the beginning of the relaxation process due to establishment of
strong correlation of structural quantities related to 4-coordinated
particles.

We observed that during initial stages of relaxation (after the R-point), 
a straight line region (SLR) is developed in the cumulative
potential energy distributions for all trajectories with an configurational
(effective) temperature close to 1060 K.  Based on study by Luedtke and 
Landman~\cite{LUEDTKE89} and also the development of correlation between
structural quantities related to 4-coordinated particles as detailed above, 
the formation
of SLR can be attributed to the freezing of the 4-coordinated amorphous network.
The distributions corresponding to
the metastable equilibrium metastable liquid phase (determined based on trajectory upto R-point)
is found to be approximately tangential to the SLR.  
Thus the SLR region with an effective temperature close to $1060$ K affects the broadness of
the metastable liquid potential energy distribution, 
when the bath temperature is in the range of 1060--1070 K.  In particular
the distribution for the metastable liquid phase broadens, as the bath temperature approaches the effective
temperature. 
This, in turn, causes increase in fluctuations of potential energy
and hence increase of heat capacity as temperature is reduced from 1070 K to 1060 K,
as observed earlier in MD simulations.~\cite{HUJO11} 

\begin{acknowledgments}
This work was supported by the young scientist scheme of the 
Department of Science and Technology, India.  
\end{acknowledgments}

\pagebreak
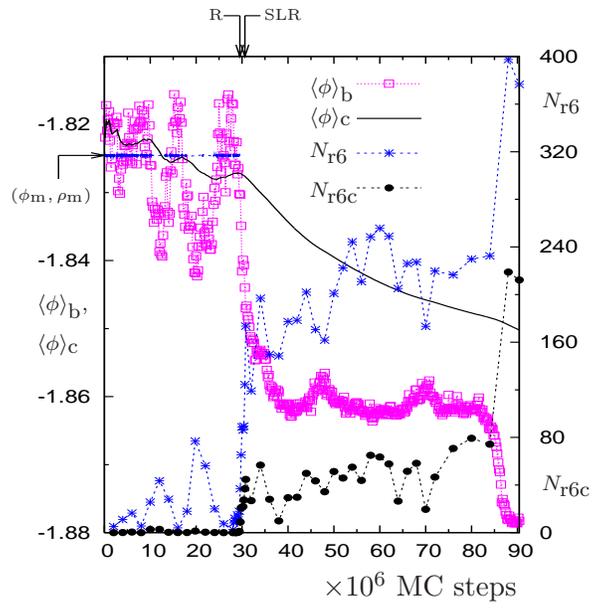
\begin{figure}
  \caption{\label{fig:t1060} 
   \normalsize{
    The NPT-MC trajectory at 1060 K  with 512 particles.     
   The trajectory is
   shown in terms of block averages obtained after every 0.2 million MC steps.  The solid black
   line shows the trajectory in terms of cumulative averages.   The black dotted line (indicated by
   horizontal arrow) denotes the points along the trajectory at which the minimum $(\phi_m, \rho_m)$ 
   of the free energy surface corresponding to the metastable liquid is accessed.  This minimum
   is located between within the rectangular area formed by the points 
   $(\phi, \rho)$ and $(\phi+\Delta \phi, \rho+\Delta \rho)$, where
   $\phi = -1.8247$, $\rho = 0.47556$, $\Delta \phi = 3 \times 10^{-4}$, and 
   $\Delta \rho = 1.4 \times 10^{-4}$.   The point along the trajectory beyond which $(\phi_m, \rho_m)$
   is not accessed is considered as the initiation point of the relaxation process and is denoted
   by `R'.  The point along the trajectories at which straight line region is formed in the
   cumulative energy distribution (see Fig.~\ref{fig:fig1}) is denoted as `SLR'.  The figure also 
   shows the number of 
   six-membered rings $N_{r6}$ and the six-membered rings with
   a chair--like structure $N_{r6c}$ (computed as block averages over 0.2 million MC
   steps) along the trajectory (please refer to ordinate on the right hand side for these quantities).
   }}
\end{figure}
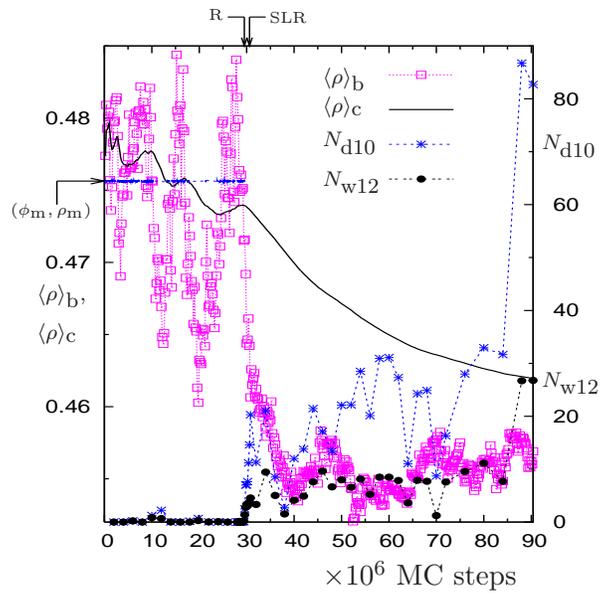
\begin{figure}
  \caption{\label{fig:str1060} 
   \normalsize{
    The structural changes for the NPT-MC trajectory at 1060 K  with 512  particles (same trajectory as
   in Fig.~\ref{fig:t1060}) along with cumulative and block averages of the density.  
    The number of diamond-like clusters with 10-atoms $N_{\mbox{d10}}$
   and wurtzite--like clusters with 12 atoms $N_{\mbox{w12}}$ (in terms of block averages taken
   over 0.2 million MC steps) are shown with reference to ordinate on the right hand side.  The
   rest of the symbols have the same meaning as in Fig.~\ref{fig:t1060}.
   }}
\end{figure}
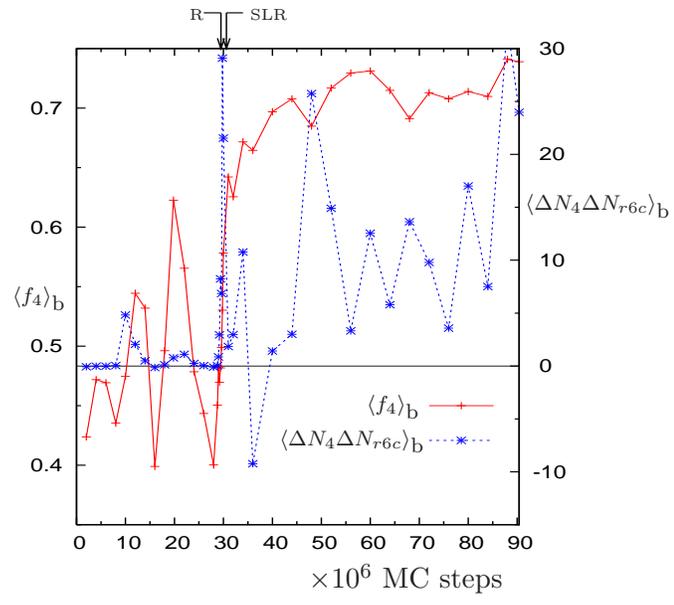
\begin{figure}
  \caption{\label{fig:n4r6c1060} 
   \normalsize{
    The block averages of the 
   fraction of 4-coordinated atoms $f_4 = N_4/N$ along the trajectory shown in 
   Figs.~\ref{fig:t1060} and     
   ~\ref{fig:str1060}, where $N_4$ is the number of 4-coordinated particles,
   and $N=512$ is the total number of particles.  The ordinate on the right
   hand side represents the correlation in the fluctuations of $N_4$ and $N_{r6c}$, 
   in terms of the block averages of the quantity 
   $\langle \Delta N_4 \Delta N_{r6c} \rangle_b$.  Here $\Delta N_4 = N_4 - \langle N_4 \rangle_b$
   is the fluctuation from the average and similarly $\Delta N_{r6c}$ is defined.
   The horizontal line (drawn as a guide to the eye)
   corresponds to $\langle \Delta N_4 \Delta N_{r6c} \rangle_b = 0$.  
   It can be seen that after the
   $R$-point, there is a sharp increase in the magnitude of the correlation.
   This shows that the transition occurs from a weakly correlated system to a strongly
   correlated system.
   }}
\end{figure}
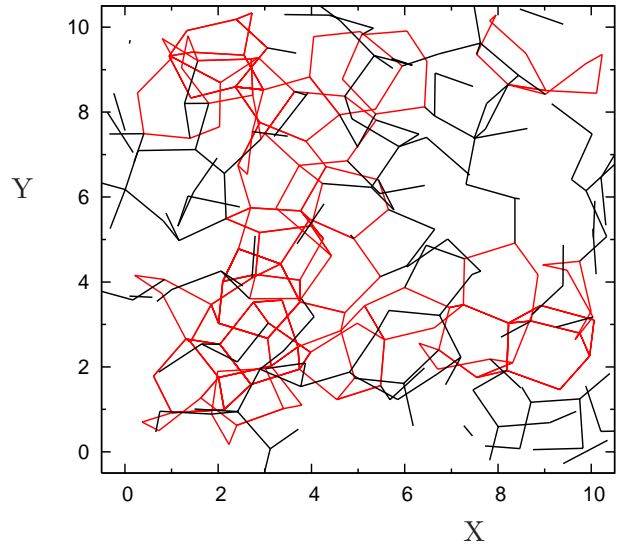
\begin{figure}
  \caption{\label{fig:config6_1} 
   \normalsize{
    The X-Y projection of 4-4 bonds, before the transition.  
      The bonds which form 6-member rings are shown in red.
    This configuration is taken at the 19.8 millionth MC step along
    the trajectory in Fig.~\ref{fig:t1060}, with a configurational
    energy of $-1.8386$ and a density of $0.4618$.
   }}
\end{figure}
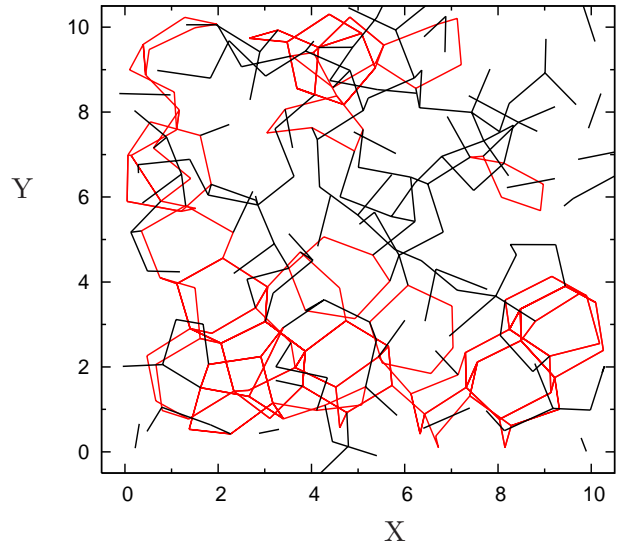
\begin{figure}
  \caption{\label{fig:config6_6} 
   \normalsize{
    The X-Y projection of 4-4 bonds, {\it after} the transition.  
     The bonds which form 6-member rings are shown in red.
    This configuration is taken at the 30 millionth MC step along
    the trajectory in Fig.~\ref{fig:t1060}, with a configurational
    energy of $-1.8429$ and a density of $0.4651$.   A crystal-like
    region formed by chair--shaped 6-member rings can be seen in 
    the lower right corner.  
   }}
\end{figure}
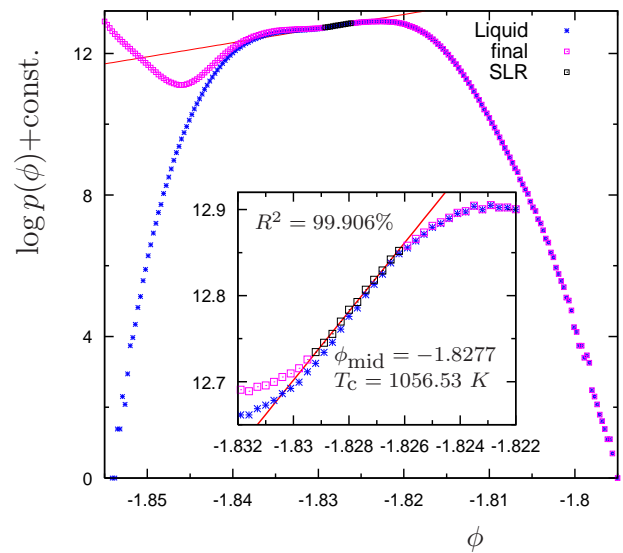
\begin{figure}
  \caption{\label{fig:fig1} 
   \normalsize{
   The cumulative potential energy distributions for the trajectory in 
   Figs.~\ref{fig:t1060} and
   ~\ref{fig:str1060}.  The ordinate is obtained as $\log N_c (\phi) = \log p(\phi) + \mbox{constant}$,
   where $N_c(\phi)$ is the number of configurations along the trajectory with per particle potential
   energy between $\phi$ and $\phi+\Delta \phi$; $\Delta \phi = 3 \times 10^{-4}$ is the
   width of the bin.  The blue stars represents the metastable liquid cumulative distribution of the 
   metastable liquid, based on 
   data collected from the trajectory upto point `R'.  The black squares in the main panel and
   the inset is the region of the distribution (called as straight line region, or SLR) 
    to which a straight line is fitted.  
     Note that SLR is completely formed at the trajectory point
    indicated in Figs.~\ref{fig:t1060} and ~\ref{fig:str1060}.
    The correlation
    coefficient, the mid point of SLR region and the configurational temperature of the SLR
    are given in the inset.  
   The `+' symbols represents {\it final} cumulative distribution obtained from trajectory upto 
   90 million MC steps.  By {\it final}, we mean that the distribution is not expected to evolve 
    further since after the system
   is unlikely to visit the part of the phase space corresponding to the range of $\phi$ values in
   the above figure.   Note that the cumulative distribution for the 
   metastable liquid phase (upto R-point),
   at least approximately, is tangential to the straight line fit to the SLR.
   }}
\end{figure}
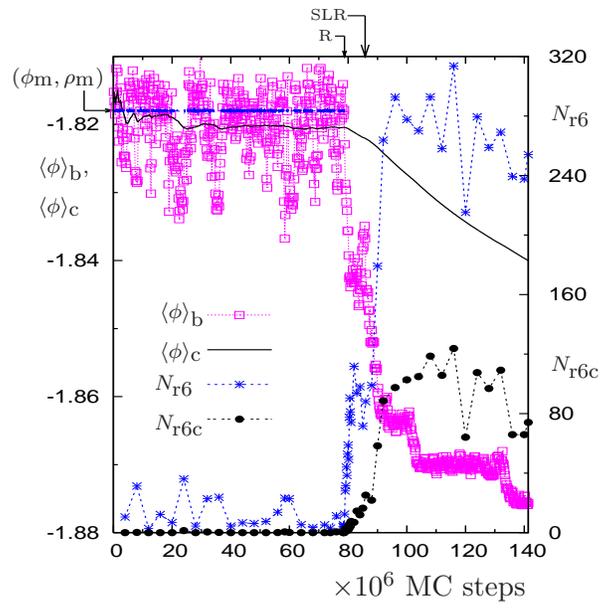
\begin{figure}
  \caption{\label{fig:t1070_N512_cp1300} 
   \normalsize{
    The NPT-MC trajectory at 1070 K  with 512 particles.     
   The trajectory is
   shown in terms of block averages obtained after every 0.2 million MC steps.  The solid black
   line shows the trajectory in terms of cumulative averages.   The black dotted line (indicated by
   horizontal arrow) denotes the points along the trajectory at which the minimum $(\phi_m, \rho_m)$ 
   of the free energy surface corresponding to the metastable liquid is accessed.  This minimum
   is located between within the rectangular area formed by the points 
   $(\phi, \rho)$ and $(\phi+\Delta \phi, \rho+\Delta \rho)$, where
   $\phi = -1.8181$, $\rho = 0.4806$, $\Delta \phi = 3 \times 10^{-4}$, and 
   $\Delta \rho = 1.4 \times 10^{-4}$.   The point along the trajectory beyond which $(\phi_m, \rho_m)$
   is not accessed is considered as the initiation point of the relaxation process and is denoted
   by `R'.  The point along the trajectories at which straight line region is formed in the
   cumulative energy distribution (see Fig.~\ref{fig:fig3}) is denoted as `SLR'.  The figure also 
   shows the number of 
   six-membered rings $N_{r6}$ and the six-membered rings with
   a chair--like structure $N_{r6c}$ (computed as block averages over 0.2 million MC
   steps) along the trajectory (please refer to ordinate on the right hand side for these quantities).
   }}
\end{figure}
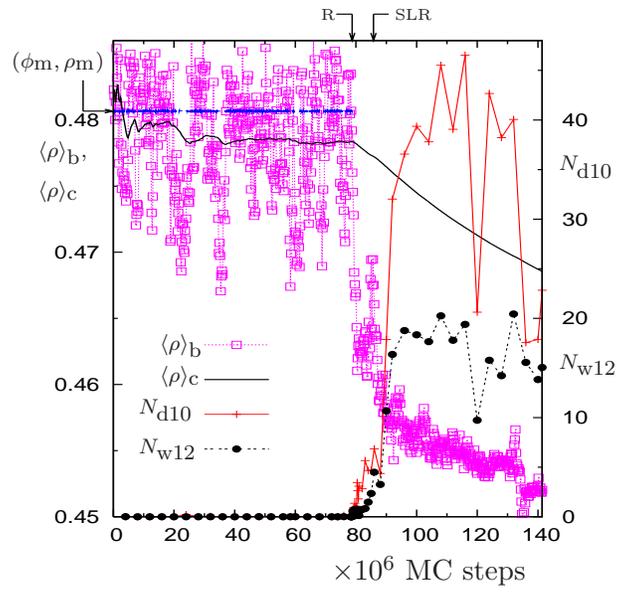
\begin{figure}
  \caption{\label{fig:str1070} 
   \normalsize{
    The structural changes for the NPT-MC trajectory at 1060 K  with 512  particles (same trajectory as
   in Fig.~\ref{fig:t1070_N512_cp1300}) along with cumulative and block averages of the density.  
    The number of diamond-like clusters with 10-atoms $N_{\mbox{d10}}$
   and wurtzite--like clusters with 12 atoms $N_{\mbox{w12}}$ (in terms of block averages taken
   over 0.2 million MC steps) are shown with reference to ordinate on the right hand side.  The
   rest of the symbols have the same meaning as in Fig.~\ref{fig:t1070_N512_cp1300}.
   }}
\end{figure}
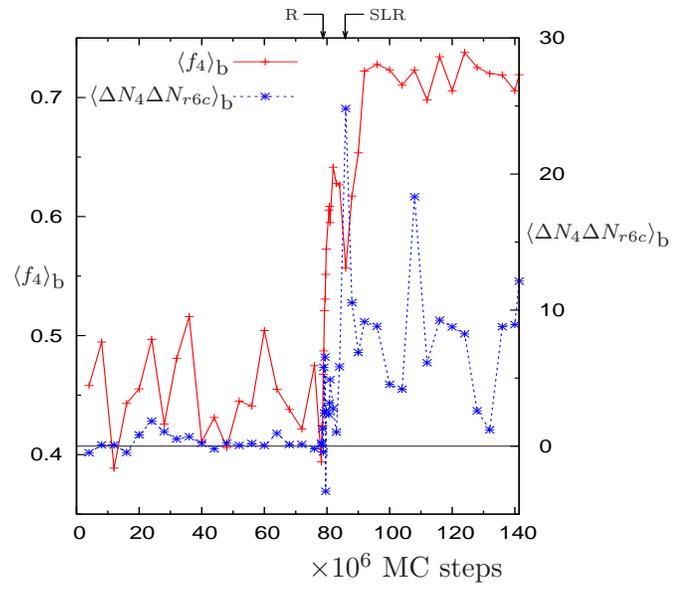
\begin{figure}
  \caption{\label{fig:n4r6c1070} 
   \normalsize{
    The block averages of the 
   fraction of 4-coordinated atoms $f_4 = N_4/N$ along the trajectory shown in 
   Figs.~\ref{fig:t1070_N512_cp1300} and     
   ~\ref{fig:str1070}, where $N_4$ is the number of 4-coordinated particles,
   and $N=512$ is the total number of particles.  The ordinate on the right
   hand side represents the correlation in the fluctuations of $N_4$ and $N_{r6c}$, 
   in terms of the block averages of the quantity 
   $\langle \Delta N_4 \Delta N_{r6c} \rangle_b$.  Here $\Delta N_4 = N_4 - \langle N_4 \rangle_b$
   is the fluctuation from the average and similarly $\Delta N_{r6c}$ is defined.
   The horizontal line (drawn as a guide to the eye)
   corresponds to $\langle \Delta N_4 \Delta N_{r6c} \rangle_b = 0$.  
   It can be seen that after the
   $R$-point, there is a sharp increase in the magnitude of the correlation.
   This shows that the transition occurs from a weakly correlated system to a strongly
   correlated system.
   }}
\end{figure}
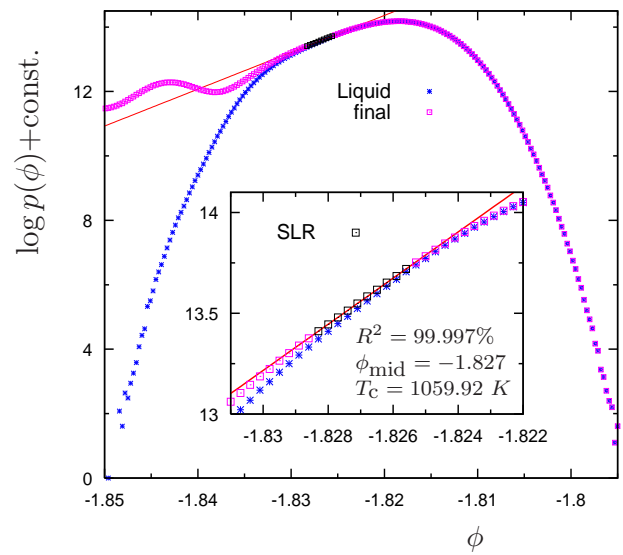
\begin{figure}
  \caption{\label{fig:fig3} 
   \normalsize{
   The cumulative potential energy distributions for the trajectory in 
   Figs.~\ref{fig:t1070_N512_cp1300} and
   ~\ref{fig:str1070}.  
    The ordinate is obtained as $\log N_c (\phi) = \log p(\phi) + \mbox{constant}$,
   where $N_c(\phi)$ is the number of configurations along the trajectory with per particle potential
   energy between $\phi$ and $\phi+\Delta \phi$; $\Delta \phi = 3 \times 10^{-4}$ is the
   width of the bin.  The blue stars represents the metastable liquid cumulative distribution, i.e., based on 
   data collected from the trajectory upto point `R'.  The black squares in the main panel and
   the inset is the region of the distribution (called as straight line region, or SLR) 
    to which a straight line is fitted.  
    The correlation
    coefficient, the mid point of SLR region and the configurational temperature of the SLR
    are given in the inset.  
    Also, note that $\phi_{\mbox{mid}}$ and $T_c$ for the SLR region is
   close to the equilibrium (average) values of metastable liquid phase, i.e., $-1.8272$ and $1060$ K, respectively
   (see Table I of Ref.~\onlinecite{APTE12}).
     Note that SLR is completely formed at the trajectory point
    indicated in Figs.~\ref{fig:t1060} and ~\ref{fig:str1060}.
   The `+' symbols represents {\it final} cumulative distribution obtained from trajectory upto 
   141.4 million MC steps.  By {\it final}, we mean that the distribution is not expected to evolve 
    further since after the system
   is unlikely to visit the part of the phase space corresponding to the range of $\phi$ values in
   the above figure.   It is also to be noted that the cumulative distribution for the 
   metastable liquid phase (upto R-point),
   at least approximately, is tangential to the straight line fit to the SLR.
   }}
\end{figure}
\end{document}